\documentclass[
  ,draft            
  ,numberedheadings 
  ]
  {aipproc}

\layoutstyle{6x9}
\def\beq{\begin{eqnarray}}
\def\eeq{\end{eqnarray}}

\newcommand{\brass}[3]{{ }^{#3}_{#2}\!\left\langle #1 \right|}
\newcommand{\ketss}[3]{\left| #1 \right\rangle_{#2}^{#3}}

\begin{document}

\title[The clock ambiguity: Implications and new developments]{The
  clock ambiguity: Implications and new developments\footnote{To
  appear in the proceedings of the {\em The Origins of Time's
  Arrow}\cite{Albrecht:2008xx}.}}  


\author{Andreas Albrecht}{
  address={University of California at Davis\\ Department of
  Physics\\ One Shields Avenue \\ Davis, CA 95616}
}

\author{Alberto Iglesias}{
  address={University of California at Davis\\ Department of
  Physics\\ One Shields Avenue \\ Davis, CA 95616}
}

\begin{abstract}
We consider the ambiguity associated with the choice of clock in time 
reparameterization invariant theories. This arbitrariness
undermines the goal of prescribing a fixed set of physical laws, since
a change of time variable can completely alter the predictions
of the theory. We review the main features of the clock ambiguity and our
earlier work on its implications for the emergence of physical laws in
a statistical manner.  We also present a number of new
results:  We show that (contrary to suggestions in our earlier work) time
independent Hamiltonians may quite generally be assumed for laws of physics that
emerge in this picture.  We also further explore the degree
to which the observed Universe can be well approximated by a random
Hamiltonian.  We discuss the possibility of predicting the dimensionality
of space, and also relate the 2nd derivative of the density of states
to the heat capacity of the
Universe. This new work adds to the viability of
our proposal that strong predictions for physical laws may emerge
based on statistical arguments despite the clock ambiguity, but many
open questions remain. 




\end{abstract}

\maketitle


\section{Introduction}
\label{sec:Intro}

Every theory that is invariant under time reparameterization presents a 
problem the moment we attempt quantization. 
Quantization gives a preferential 
role to time (in the definition of canonical variables) that cannot be 
fulfilled in a theory that is unaltered by its reparameterization. 
A prominent example of such a theory is given by General Relativity 
and in this context 
there have been extensive discussions of the problem 
(see, for example, 
\cite{Arnowitt:1960es} for an early treatment or \cite{Isham:1992ms} for an
comprehensive review). An approach often used in cosmology is to work in
``superspace'' finding time as an ``internal'' variable after quantization.
The invariance is imposed on the quantum states of the superspace 
$|\psi\rangle_S$ as a physical condition involving the Hamiltonian constraint,
\beq
{\cal H}|\psi\rangle_S=0~.
\eeq  
In \cite{Albrecht:1994bg,Albrecht:2007mm}, 
we argued that such an approach carries an intrinsic arbitrariness in the 
choice of ``clock'' subspace that leads in turn to an arbitrariness in the 
predictions of the theory; the clock ambiguity. We showed that its implications
are so severe that we may need to see the laws of physics as we know them as 
an approximate emergent phenomenon.  

By taking the clock ambiguity seriously, we look for the emergence of physical 
properties derived from a Hamiltonian evolution chosen randomly, corresponding 
to an absolute ambiguity in the choice of clock.  In
\cite{Albrecht:2007mm} we singled out quasiseparability as a crucial
feature of physical laws needed to 
sustain observers, and argued that quasiseparability is optimally
achieved through locality (and thus through local field theory).  In that
context, we find our result from \cite{Albrecht:2007mm} that any sufficiently 
large random Hamiltonian can be interpreted (to a sufficiently good
approximation) as a local field theory encouraging: It suggests that
combining the randomness suggested by the clock ambiguity with the
need for quasiseparability could yield local field theory as a {\em
prediction}.

In this work, section \ref{sec:Summary} reviews the clock ambiguity and 
sketches the basic approach we advocated in \cite{Albrecht:2007mm} to seek 
predictive power based on a statistical analysis.  Section 
\ref{sec:TimInd} gives a new result that shows that one can quite generally
take the physical laws that emerge in our analysis to have a time {\em
independent} Hamiltonian (this result is in contrast to assumptions
we made in our earlier work). Section \ref{sec:FTWig} reviews our analysis
from \cite{Albrecht:2007mm} showing that any sufficiently large
random Hamiltonian can be interpreted, to a good approximation, as a
local field theory.  In section \ref{sec:FTWigDim} we extend that work
to discuss the possibility of predicting the dimensionality of space,
and apply our analysis to a non-standard distribution of random
Hamiltonians in section \ref{sec:FTWigTail}, with interesting
implications for higher orders in our Taylor series comparison of
random Hamiltonians with field theories. After reviewing our thinking
about gravity in this picture in section \ref{sec:Gravity}, we extend
our treatment to gravitating systems in section \ref{sec:Heat} by relating the
derivatives of the density of states to appropriate thermodynamic
quantities which can be estimated for gravitating systems.  The result of this extension, while very crude,
is encouraging.  We present our conclusions in section
\ref{sec:Conclusions}

\section{Summary of the clock ambiguity}
\label{sec:Summary}

The clock ambiguity arises from the treatment of time as ``internal''
in time reparameterization invariant theories. ``Internal time''
means that a subsystem of the universe is identified as the time
parameter or ``clock''
and time evolution is revealed by examining correlations between the
rest of the universe and the clock subsystem.  In quantum theories
this picture is typically expressed in ``superspace'', of which the
clock system is a subspace.

In previous work \cite{Albrecht:1994bg,Albrecht:2007mm} we pointed out
that regardless of how careful one is to describe a universe as obeying
specific physical laws, the same state in the same superspace can equally
well describe a completely different physical world with completely
different time evolution.  One only has to identity a different clock
subsystem to find this new description.   This is the clock
ambiguity.  We have shown that the clock ambiguity is absolute, in the
sense that all possible systems experiencing all possible time
evolution can be extracted from the same superspace state by a suitable 
choice of clock. 

We refer the reader to this earlier work for the
details \cite{Albrecht:1994bg,Albrecht:2007mm}.  Here we quote the main
result.  We assume a discrete formalism which allows us to write 
the state in superspace as
\begin{equation}
\ketss{\psi}{S}{} =
\sum_{ij}\alpha_{ij}\ketss{t_i}{C}{}\ketss{j}{R}{} \equiv
\sum_i \ketss{t_i}{C}{}\ketss{\phi_i}{R}{}.
\label{CRexpand}
\end{equation}
Here the subscripts $S$, $C$ and $R$ relate to the decomposition of superspace 
$S$ according to $S = C\otimes R$, and refer to the full superspace, the
clock subspace and the ``rest'' of the superspace respectively.
The bases $\ketss{t_i}{C}{}$ and $\ketss{j}{R}{}$ span the clock and
``rest'' subspaces.  The second equality defines (by summing over $j$)
$\ketss{\phi_i}{R}{}$, giving the wavefunctions of the ``rest'' subspace at 
times $t_i$.

One can see that all the information about the state in the $R$
subspace  and its
time evolution is contained within the expansion coefficients
$\alpha_{ij}$. In \cite{Albrecht:1994bg,Albrecht:2007mm} we show that
arbitrary values $\alpha^{\prime}_{ij}$ can result from expressing the
same superspace state $\ketss{\psi}{S}{}$ according to suitable
choices of the decomposition $S=C^{\prime}\otimes R^{\prime}$, or in
other words, by making a suitable choice of clock.  Thus any state
evolving according to any Hamiltonian can be found, merely by choosing
a new clock in the superspace.  

One possible conclusion from the clock ambiguity is that the formalism
that leads to this result must be wrong in some way (that
in itself would have interesting implications).  Otherwise, if
we conclude that our fundamental theories really must have the clock 
ambiguity, the success of physics so far implies that it must be possible 
to come up with sharp predictions of specific physical laws, presumably 
based on some kind of statistical arguments, given that all possible 
physical laws are represented in the formalism\footnote{
We have recently learned that Chris Wetterich has considered very
similar issues using the functional integral formalism\cite{Wetterich:1988uq,Wetterich:1992uz}
.}.

In \cite{Albrecht:2007mm} we explored how one might go about formulating
such a statistical analysis, and gave special emphasis to the
quasiseparability of physical laws which seems so curial for our
ability to survive and thrive as tiny observers.  We noted that
locality (as realized in the local field theories that describe the
elementary particles and forces) is the ultimate origin of the 
quasiseparability we experience in our physical world.  We also noted that 
in some sense local field theories give a maximal expression of quasi-locality. 
Thus we feel our result from \cite{Albrecht:2007mm}, that any random
Hamiltonian can yield a sufficiently good approximation to a local
field theory is quite interesting. It suggests that the requirement of 
quasiseparability may universally lead to local field theories as one
searches for emergent physical laws in theories with the clock
ambiguity. We review and extend that result in section \ref{sec:FTWig}. 

\section{The time independence of $H$}
\label{sec:TimInd}

A randomly chosen clock leads to a randomly chosen set of
$\alpha_{ij}$'s. Random $\alpha_{ij}$'s describe a randomly chosen
state evolving under a random Hamiltonian.  The lack of any a-priori
reason to expect correlations between the $\alpha_{ij}$'s with
different $i$ values suggests that in general
the random Hamiltonian will be different for each time step (labeled
by $i$). We discuss this issue in section III-B of
\cite{Albrecht:2007mm}.

However, in our earlier work we overlooked a rather simple point
(kindly brought to our attention by Glenn
Starkman \cite{Starkman:2007aa}).  The point is that $\alpha_{ij}$'s do
not contain nearly enough information to specify a full Hamiltonian at
each time.  We can use this fact to add a requirement that the Hamiltonian
is time independent without any loss of generality, assuming one does
not take too many time steps.  We show below that this constraint is
very easy to meet. 

A time step can be written as 
\begin{equation}
  \ketss{\psi(t_{i+1})}{R}{} = \left[{\bf 1}-i\hbar(\Delta t_i){\bf
  H}(t_i)\right]\ketss{\psi(t_i)}{R}{}.
\label{Hdef}
\end{equation}
By taking the inner product of this equation with  $\brass{\psi(t_{i+1})}{R}{}$ 
one finds
\begin{equation}
1 =  {}_R\langle \psi \left(
  {t_{i + 1} } \right) |{\bf 1} - i\hbar\left( {\Delta t_i} \right) {\bf H}\left( t_i \right) |
\psi \left( t_i \right) \rangle_R.  
\label{dt1}
\end{equation}
The inner product with $\brass{\psi^\perp (t_{i+1})}{R}{}$ gives
\begin{equation}
0 =  {}_R\langle \psi^\perp
  \left(t_{i+1}\right)\left| {\bf 1} - i\hbar\left( {\Delta t_i} \right){\bf H}\left(t_i\right)
\right| \psi \left(t_i\right)\rangle_R  
\label{dt0}
\end{equation}
where $\brass{\psi^\perp (t_{i+1})}{R}{}$ could be any one of $N-1$
states orthogonal to $\brass{\psi(t_{i+1})}{R}{}$. As shown in
Eqn. \ref{CRexpand}, the $\alpha_{ij}$ lead directly to the time
evolving state vector $\ketss{\psi(t_i)}{R}{}$. One 
uses the information from the state vector at each time step to infer
information about $H$.  Together Eqns. \ref{dt1} and \ref{dt0} give a 
total of $N_R$ complex (or $2N_R$ real) constraints on $H$. 
Since a general $N_R\times N_R$ Hamiltonian has $N_H^2$ real degrees
of freedom, the $\alpha_{ij}$'s do not contain
enough information to define a full Hamiltonian at each time step.  After all, 
the $\alpha_{ij}$'s only tell us about the evolution of a single state,
whereas the Hamiltonian contains full information about the evolution
of all possible states. 

The fact that the Hamiltonian is highly underdetermined by a single
time step can be exploited to add the condition that
the Hamiltonian is time independent without loss of generality.  As
long as one is looking at no more than $N_H/2$ time steps,
Eqns. \ref{dt1} and \ref{dt0} provide no more than $N_H^2$ real
constraints which can be used to build at least one time independent
Hamiltonian that describes the full time evolution.  And to the extent
that the $\alpha_{ij}$ are randomly generated, the Hamiltonians
produced from the $\alpha_{ij}$'s should be randomly distributed as
well.  In fact, it seems reasonable to expect that the central limit theorem 
will give the distribution of Hamiltonians (generated by effectively inverting
Eqns. \ref{dt1} and \ref{dt0}) an enhanced degree of Gaussianity over whatever
distribution generated the $\alpha_{ij}$'s. 

For all this to work out, we need to constrain the number of time
steps $N_t$ according to 
\begin{equation}
  N_t < N_H/2.
\label{ntbound}
\end{equation}
We can estimate $N_t$ as the age of the Universe divided by the minimum
time resolution $\delta t$.  Using arguments from section \ref{sec:FTWig}, 
$\delta t \equiv 1/\Delta E$ and the maximum value of $\Delta E$ 
($=10^{11}GeV$)  gives 
\begin{equation}
  N_t \approx \frac{\Delta E}{H_0} = \frac{10^{11}GeV}{10^{-42}GeV} = 10^{53}.
\label{ntboundN}
\end{equation}
By comparison, requiring a good match of the density of states to a
field theory leads to Eqn. \ref{m1} giving
\begin{equation}
  N_H \geq \frac{B}{a}\frac{E_M }{E_0 }
\left[1-\left( \frac{E_0-E_S}{E_M}\right)^\beta\right]^{-\gamma}
\!\!\exp\left[b\left(\frac{E_0}{\Delta k}\right)^\alpha\right]
\label{eqn:Zeq}
\end{equation}
The quantity $E_0/\Delta k$ in the exponent is the ratio of the energy
of the Universe to the field theory $k$-space cutoff. Even choosing values from
section \ref{sec:FTWig} which minimize the bound on $N_H$ give
exponentially large values for the {\em exponent} in
Eqn. \ref{eqn:Zeq} and give lower bounds on $N_H$ which easily satisfy
Eqn. \ref{ntbound} and validate the assumption of a time independent
Hamiltonian\footnote{This argument appears to be very robust. For
  example, refining the time resolution to $\delta t =1/M_P$ does not
  change the result at all.}

\section{Field theory and the Wigner semicircle}
\label{sec:FTWig}

\subsection{Our basic approach}
\label{sec:FTWigBasic}

The clock ambiguity implies that any random split of superspace into
clock and rest subsystems should lead to a realization of ``physical laws''. 
However, one expects that a random
split would result in laws described by a random Hamiltonian. In 
\cite{Albrecht:2007mm} we discussed possible ways forward under
those conditions.  One thing we did was pose the question in the
converse form to test this hypothesis. Namely, we  
evaluated the extent to which the known physical laws match to those derived 
from a random Hamiltonian evolution. 
In particular, we compared the spectrum of a
free field theory, representing (approximately) the known physics, to the 
eigenvalue spectrum of random Hamiltonians.  

Following \cite{Albrecht:2007mm}, we do not undertake the project of
specifically constructing field operators etc. in terms of the
eigenstates of the Hamiltonian.  This project is likely to be
challenging, and is also likely to further involve a statistical
analysis of different physical realizations consistent with the same
eigenvalue spectrum and initial state $\ketss{\psi (t_1)}{R}{}$.  We
feel that our analysis at the level of the eigenvalue spectrum
represents a first check of the viability of our line of reasoning,
and we save the important question of defining field operators etc. for
future stages of this work\footnote{When we presented this work at the
 {\it Origins of Time's Arrow} meeting Lee Smolin drew our attention
 to work by Bennett {\em et al.} \cite{Bennett:1987xx} which may offer
 a framework where specific symmetries and representations for
 elementary particles could be predicted in a scheme such as ours.}

The distribution of eigenvalues for a random Hamiltonian, represented as an
$N_H\times N_H$ Hermitian matrix,
follows under quite general assumptions \cite{Mehta} the Wigner
semicircle rule in the large $N_H$ limit. Take, for example, the
distribution of eigenvalues of a large Hermitian matrix with elements
drawn from a Gaussian distribution depicted in Fig. \ref{fig:ws}.

On the other hand, the density of states for a free field
theory grows, at large energies, like an exponential of a 
power of the energy.  On the face of it, these two forms for $dN/dE$
are dramatically different. In order to press forward with the comparison we 
introduced a general parametrization for the random Hamiltonian and field theory
spectral densities respectively:
\begin{eqnarray}\label{dr}
{dN_R\over dE}&=&\left\{
\begin{array}{ll}
a{N_H\over E_M}\left(1-\left({E-E_S\over E_M}\right)^\beta\right)^\gamma & 
|E-E_S|<E_M~,
\\
0 & {\rm otherwise}~, \end{array}
\right.\\
{dN_F\over dE}&=&{B\over E}\exp\left\{b\left(E\over\Delta
k\right)^\alpha\right\}~,\label{df}
\end{eqnarray}
where $E_M$ and $E_S$ represent the maximum eigenvalue of the random
Hamiltonian and an offset
energy between the two descriptions, $\Delta k$ ($\equiv 2\pi/L$) is the 
resolution in $k$-space set by putting the field theory in a box of size $L$ 
and $B$, $b$, $\alpha$ and $\gamma$ are dimensionless parameters.

Expanding both Eqns. \ref{dr} and \ref{df} in a Taylor series
around a given central energy $E_0=\rho R_H^3=10^{80}GeV$, corresponding to 
the current energy
of the Universe, and trying to equate the results at each order in 
$(E-E_0)$ we find the level of agreement between the two descriptions.

Equating the zeroth order sets the size of the space of the random Hamiltonian 
to be exponentially large:
\beq\label{m1}
N_H = {B\over a}{E_M\over E_0}
\left[1-\left(E_0-E_S\over E_M\right)^\beta\right]^{-\gamma}
\exp \left[b\left(E_0\over \Delta k\right)^\alpha\right]~.
\eeq
Strictly speaking, this expression only gives a lower bound on $N_H$, since we 
only really know upper bounds on $\Delta k$.  

Equating the first order (as well assuming
equality at zeroth order) sets the offset 
energy $E_S$ in terms of the energy of the Universe $E_0$ by the
following implicit expression:
\beq\label{m2}
-\beta\gamma{E_0\over E_0-E_S}
{\left(E_0-E_S\over E_M\right)^\beta\over 1-\left(E_0-E_S\over E_M\right)^\beta}
=\alpha b \left(E_0\over \Delta k\right)^\alpha~.
\eeq
\begin{figure}
\includegraphics[width=3.5in]{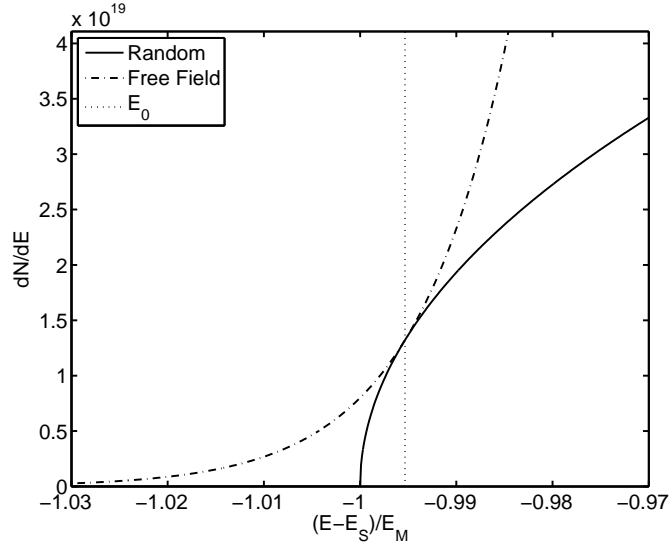}
\caption{\label{fig:both}
A plot of the density of eigenvalues for a random Hamiltonian using
Eqn.~\ref{dr} and a field theory using Eqn. \ref{df} matching the zeroth and 
first order terms in a Taylor expansion around $E_0$ (the vertical line).  
}
\end{figure}

Assuming equality and 0th and 1st order, the relative difference at
the second order is fixed and given by
\begin{eqnarray}\label{delta2}
\Delta_2\equiv {\Delta {dN\over dE}\over {dN\over dE}|_{E_0}}\approx
{\alpha^2 b^2\over\gamma}
\left(E_0\over \Delta k\right)^{2\alpha} {(E-E_0)^2\over E_0^2}~.
\end{eqnarray}
Table \ref{table} shows the value of $\Delta_2$ for different values of the 
exponent $\alpha$ in Eqn. \ref{df}, 
the  field theory $k$-space lattice spacing $\Delta k$ and 
the range of validity of the field theoretical description
\begin{equation}
\Delta E=E-E_0  
\label{DeltaE}
\end{equation}
which can be thought of in terms of a minimum timescale on which field theory is
valid, given by $\delta t \sim 1/\Delta E$.  The idea is to check if the 
disagreement between the density of states
of a random Hamiltonian and a free field theory  at 2nd order, $\Delta_2$ can
be ``sufficiently small'' for realistic parameters. We find that the
parameter most critical to this analysis is $\alpha$, and we discuss
its value in the next section.

\begin{table}
\begin{tabular}{llll}
\hline
\tablehead{1}{c}{b}{$\alpha$} & 
\tablehead{1}{r}{b}{$\Delta k$(GeV)} &
\tablehead{1}{r}{b}{$\Delta E$(GeV)} & 
\tablehead{1}{c}{b}{$\Delta_2$} \\ 
\hline
$1/2$  & $10^{-25}$  & $10^{3}$ & $10^{-24.5}$ \\
$1/2$  & $10^{-25}$  & $10^{11}$ & $10^{-16.5}$ \\
$1/2$  & $10^{-42}$  & $10^3$ & $10^{-16}$ \\
$1/2$  & $10^{-42}$  & $10^{11}$ & $10^{-8}$ \\
$3/4$  & $10^{-25}$  & $10^{3}$ & $10^{1.8}$ \\
$3/4$  & $10^{-25}$  & $10^{11}$ & $10^{9.8}$ \\
$3/4$  & $10^{-42}$  & $10^3$ & $10^{14.5}$ \\
$3/4$  & $10^{-42}$  & $10^{11}$ & $10^{22.5}$ \\
$1$  & $10^{-25}$  & $10^{3}$ & $10^{28}$ \\
$1$  & $10^{-25}$  & $10^{11}$ & $10^{36}$ \\
$1$  & $10^{-42}$  & $10^3$ & $10^{45}$ \\
$1$  & $10^{-42}$  & $10^{11}$ & $10^{53}$ \\ \hline
\end{tabular}
\caption{\label{table}
Value of $\Delta_2$ for different choices of 
$\alpha$, $\Delta k$ and $\Delta E$. As in
\cite{Albrecht:2007mm}, values for $\Delta E$ are set by accelerator
($10^{3}GeV$) or cosmic ray ($10^{11}GeV$) bounds. Values for $\Delta k$
are set by the photon mass bound ($10^{-25}GeV$) or the size of the
Universe ($10^{-42}GeV$).
} 
\end{table}

\subsection{The value of $\alpha$ and the dimensions of space}
\label{sec:FTWigDim}

The results for the density of states of a field theory in
$1+1$ dimensions for bosons and fermions can be derived from different
instances of the Cardy formula for conformal field theories in 2d 
\cite{Cardy:1986ie}.
This formula relates the entropy of the field theory to its energy $E$ and
central charge $c$ in the following way
\begin{eqnarray}
S=\log N(E)={1\over 2\pi}\sqrt{{c\over 6}(E-{c\over 24})}~,
\end{eqnarray}
and implies Eqn. \ref{df} with exponent $\alpha=1/2$.  
The asymptotic density
of states can also be found for a conformal field theory in higher number
of dimensions \cite{Banks:1999az} and grows as
$e^{E_E^{(d-1)/ d}}$ where $E_E$ is the extensive energy. However,
if the Casimir energy $E_C$ is taken into account the total energy $E=E_E+E_C$
is sub-extensive and the dependence of the entropy on energy changes.
Verlinde \cite{Verlinde:2000wg}, based on holographic arguments, proposed 
that the Cardy formula is satisfied also in the case of higher dimensional 
field theories.
  
Taking the extensive energy expression for a field theory in $3+1$ dimensions 
would fix the constant $\alpha=3/4$ in our parametrization of the density of 
states Eqn. \ref{df}. A first assessment of table \ref{table} indicates
that the agreement between the field theory and random Hamiltonian 
would be poor (with $\alpha=3/4$, $\Delta_2\gg 1$
for all entries). An alternative interpretation might be to note that
the transition from $\alpha=1/2$ to $\alpha=3/4$ in our table occurs
roughly right at the point where $\Delta_2$ shifts from small to large
values. Given that all our estimates are very rough at this stage,
there may be a hint here of a way in which our methods could {\em predict} the
three dimensions of space, as the maximum value consistent with a
random Hamiltonian.

On the other hand, if we assume Verlinde is correct and use the
universal Cardy formula, that implies $\alpha=1/2$ for
any $d$.  Then the difference $\Delta_2$ is negligible and random Hamiltonians 
give a density of states that appears strongly consistent with the
field theoretical one, at the expense of any apparent preference for
the value of $d$.



\subsection{Wigner's tail}
\label{sec:FTWigTail}

It may appear disturbing that we are attempting to match expressions  
Eqns. \ref{dr} and \ref{df}, the latter having positive second derivative 
everywhere
while the former in the case of the Wigner semicircle is negative definite; the
case depicted in Fig. \ref{fig:both}. As discussed above, it may
simply be the case that this discrepancy is negligible, and is not a
problem.  

\begin{figure}
\includegraphics[width=4in]{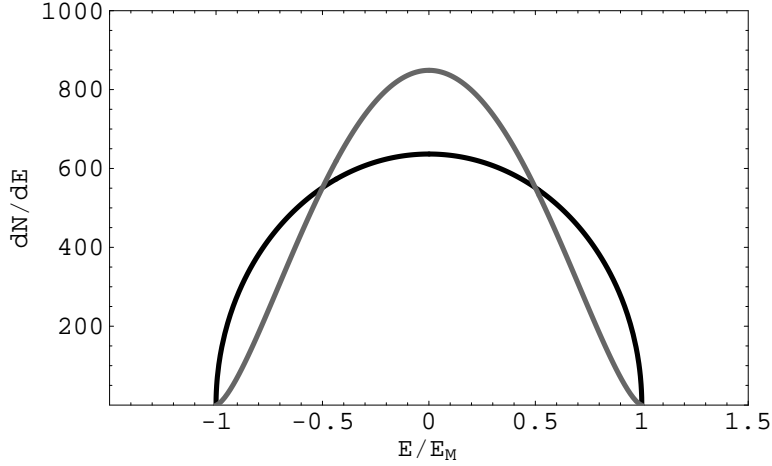}
\caption{\label{fig:ws}
A plot of the density of eigenvalues for a random Hamiltonian 
($E_M=1$, $N_H=1000$) in the cases of: a Gaussian distribution (black) 
giving rise to the Wigner Semicircle and 
Gaussian plus a cubic ``interaction'' term (gray) with concave tails. 
}
\end{figure}

One might also wonder if this may change if the
perfectly Gaussian probability distribution is altered, for example, if
the width of the distribution of eigenvalues is different in different energy
ranges\footnote{We thank Jaume Garriga for suggesting this direction
  of  investigation.}. 
To be concrete, one may consider the distribution containing a  
small cubic piece. In such a case (studied in \cite{Brezin:1977sv}) the
exponent $\gamma$ in the density of states may be changed from $1/2$ (Wigner
semicircle) to $3/2$ which has regions of positive second derivative near
the tails of the distribution as depicted in Fig. \ref{fig:ws}. 
This possibility is included in
our parametrization given in Eqn. \ref{dr}. The corresponding improvement in 
matching can be inferred from Eqn. \ref{delta2}; an increase in $\gamma$ leads 
to a smaller relative difference $\Delta_2$. 

Let us point out, as a curiosity, that a distribution highly distorted from 
Gaussianity might lead to a perfect matching with the field theory 
distribution.
In fact, letting $\gamma$ grow makes the 
generalized random density of states (Eqn. \ref{dr}) approach an
exponential of the  
form of the field theory one (Eqn. \ref{df}). In order to see this we may take 
$E_M\gg E_0-E_S$ in Eqn. \ref{m2} to find 
\beq\label{m2appr}
-\gamma\left(E_0-E_S\over E_M\right)^\beta\approx\left[{\alpha\over\beta}
\left(1-{E_S\over E_0}\right)\right]b\left({E_0\over\Delta k}\right)^\alpha~,
\eeq
and choose parameters such that the coefficient in brackets is approximately 
one.
Therefore, we have that the random density of states has the form 
\beq
{dN_R\over dE}=a {N\over E_M}\left(1+{x\over\gamma}\right)^\gamma~,
\eeq
where,
$x\equiv-\gamma(E-E_S/E_M)^\beta\approx b(E/\Delta k)^\alpha$ for 
$E\approx E_0$, that in the limit 
of large $\gamma$ reproduces the exponential behavior of the field theory 
density of states. However, we don't think that such a distortion of the 
distribution could be the outcome of a truly statistical averaging
procedure.  Furthermore, it seems contradictory to the spirit of this
work to seek out an exotic distribution.  That would appear to undermine
the hope that the our methods could one day offer some real predictive
power. 

\section{Including Gravity}
\label{sec:Gravity}
In this and previous work we have not discussed gravity at length. In
\cite{Albrecht:2007mm} we suggested that gravity could naturally emerge 
when a more general metric is allowed when interpreting a
random Hamiltonian as a local field theory (vs. the Minkowski metric
implicit in the discussion in section \ref{sec:FTWig}). In such a picture we
do not expect a full consistent theory describing arbitrary spacetimes
to emerge.  It would be enough to get a theory of spacetime that would
be consistent for the actual state of the Universe and similar
states. It is not even clear, for example, that the full number of
states associated with black hole entropy would need to be part of the
spectrum in such a picture, since the microscopic properties of black
holes do not really appear to be part of our physical world .  It seems 
reasonable to proceed carefully with this issue, and avoid jumping to 
conclusions about gravity in this picture until some of these ideas have 
been worked out more systematically\footnote{We find it intriguing that this 
picture bears some resemblance to approaches that explicitly reject a full 
``third quantized'' superspace formalism, such as that discussed in 
\cite{Banks:2002wr}.}

In the next section we will try a different approach.  Specifically,
we will relate the curvature of the Wigner semicircle to the
specific heat of the Universe. In estimating the specific heat we use
standard notions of the heat capacity of gravitating systems, and
thereby implicitly introduce gravity into our analysis. We do this with
the caveat that this approach may take us even further out on a limb
than the other (admittedly speculative) ideas discussed elsewhere in
this paper. Interestingly, the analysis in the next section yields
intriguing results even when the more exotic forms of gravitational
entropy (black hole and De Sitter entropy) are ignored.  Thus the
analysis of Section \ref{sec:Heat} seems to apply even in the context
of the more conservative ideas about gravity reviewed in this section.

\section{Heat capacity and $N^{\prime\prime\prime}$}
\label{sec:Heat}

Here we return to curvature of the $dN/dE$ vs $E$ curve, {\it i.e.,}
the third derivative of $N(E)$, and estimate its value from a thermodynamic
perspective. We will use the fact that the heat capacity (or its intensive 
counterpart, the specific heat) is a thermodynamic quantity related to 
$N^{\prime\prime\prime}$. As discussed in the previous section, we will 
incorporate gravity by considering thermodynamic quantities defined for 
gravitating systems such as black holes\footnote{This section
  differs significantly from the 1st version posted on the ArXiv.
  An error in Eqn. 25 of V1 (which is clearly dimensionally wrong)
  propagated to a number of places in that section.  In this version
  we have corrected the error and subsequent discussion.  
}.  

Our starting point is the standard canonical ensemble expression for
the entropy of a system with energy in a range $\Delta E$ around a
central energy $E_0$:
\begin{equation}
  S = \log\left({dN(E_0)\over dE}\Delta E\right)
\end{equation}
This leads to 
\beq
{1\over T} \equiv {dS\over dE}={d(\log({dN\over dE}\Delta E))\over dE}=
{ N^{\prime\prime}\over N^\prime}~,
\label{Tdef}
\eeq
and using $C \equiv dE/dT$
\beq\label{C}
{1\over C}={d\over dE}\left({N^\prime\over N^{\prime\prime}}\right)
=1-{N^\prime N^{\prime\prime\prime}\over N^{\prime\prime~2}}~.
\label{InvC}
\eeq
When discussing these thermodynamic quantities one must generally be
careful to state what is being varied and what is being held fixed when
differentiating.  We will return to that question a bit later in this
section. 

Plugging the generalized Wigner form (Eqn. \ref{dr}) for the density
of states into Eqn. \ref{InvC} gives
\beq
{1\over C}=1-{(\gamma-1)Q-(\beta-1)\over 2\beta^2\gamma Q}~.
\eeq
Here $1/Q=((E_0-E_S)/E_M)^\beta-1$ is an exponentially small quantity 
if the order by order matching described in section 
\ref{sec:FTWigBasic} is performed. Thus, to an excellent approximation we have
\beq
C=\left(1+{1-\gamma\over 2\beta^2\gamma}\right)^{-1}~.
\eeq  
 Taking parameters around the Winger case ($\gamma = 1/2$ and
$\beta = 2$) gives $C=9/8$.  

Originally, the motivation for this line of investigation was the
following: The second derivative of the density of states of the Wigner
semicircle has the opposite sign to that of the field theory density
of states (as can be seen by inspecting Fig. \ref{fig:both}).  The heat
capacity is related to the second derivative of the density of states,
and is negative for strongly gravitating systems.  Strongly
gravitating systems dominate the entropy of the Universe, so perhaps the negative
specific heat of strongly gravitating systems in the Universe allows
one to more fully reconcile the density of states of real matter with the Wigner
semicircle at second order.  This idea is not realized however,
because the the second derivative of the density of states is not
related to the specific heat in a sufficiently simple way.  For the
cases we consider, the second derivative of the density of states remains
positive, even when the specific heat is negative.  Forced to abandon
this simple idea, we none the less move forward with the comparison with
thermodynamic quantities which still turns out to give interesting
results. 

Due to the additivity of the entropy, it will be convenient to work
with the derivatives of entropy with respect to $E$
\beq\label{Sn}
d^nS/dE^n.
\eeq
These quantities can be constructed by differentiating $S(E)$ directly, or
they can be constructed from other thermodynamic quantities.  For example, the $n=2$ case can be related to the heat
capacity using 
\beq\label{t2c}
{1\over T^2C}=-{d^2S\over dE^2}
\eeq
(which can be derived from Eqns. \ref{Tdef} and \ref{InvC}). 

If we
write the entropy of the Universe as a sum over different 
components (such as radiation and black holes) labeled by $i$ one has
\beq\label{t2cS}
{d^2S_{tot}\over dE^2} = {d^2\over dE^2}\sum_i S_i = \sum_i S^{\prime\prime}_i 
= -\sum_i {1\over T_i^2C_i}~.
\eeq
The Wigner density of states (Eqn. \ref{dr}) gives
\beq\label{t2cR}
-{d^2S\over dE^2}=
{N^{\prime\prime~2}-N^\prime N^{\prime\prime\prime}\over N^{\prime~2}}\
=\frac{1+\beta Q}{E-E_S}{dS\over dE}=\gamma\beta\frac{(1+Q)(1+\beta Q)}
{(E-E_S)^2}~. 
\eeq

We wish to compare Eqn. \ref{t2cR} with Eqn. \ref{t2cS}.  To do so we
will either estimate $T_i$ and $C_i$ or $S_i^{\prime\prime}$ directly for
the various components of the Universe. We consider four main 
contributions coming from radiation ($R$), black holes ($BH$), dark
matter ($DM$) and dark energy ($DE$).  

{\bf Radiation:} To compute the radiation component we take a gas of photons 
with energy
$E_R=\rho_RH^{-3}=T_R^4H^{-3}$ and temperature $T_R=10^{-13}GeV$. 
Keeping the volume $H^{-3}=(10^{-42}GeV)^{-3}$ fixed we obtain
\beq\label{cr}
C_R=2\times 10^{88}~,~~~~~~~{1\over T_R^2C_R}=10^{-62}GeV^{-2}~,
\eeq
and entropy $S_R=4C_R/3\sim 10^{88}$.

{\bf Black Holes:} We use the total black 
hole entropy estimate of \cite{Kephart:2002bf}: 
\beq
S^{tot}_{BH}=\sum_{N_{gal}}4\pi {M^2_{BH}(gal)\over m_{pl}^2}\sim 
3.2\times 10^{101}{E_{BH}\over 10^{75}GeV}\left(M\over 10^{7}M_\odot\right)~,
\eeq
where the sum is over galaxies ($N_{gal}\sim 10^{11}$) within the volume 
$H^{-3}$ and $M_{BH}(gal)$ is 
the distribution of masses of supermassive black holes at the galactic 
cores, which we approximate here as being peaked at 
$M = 10^{7}M_{\odot}$.  Using
$E_{BH}=N_{gal}M=10^{11}10^7M_\odot=10^{-5}E_0$ and
$M_\odot=10^{57}GeV$ ({\it i.e.}, $T_{BH}\sim 10^{64}GeV$ ) we obtain 
\beq
\label{cbh}
C_{BH}=-2.1\times 10^{91}\left(M\over 10^{7}M_{\odot}\right)^2~,~~~~~~~
{1\over T_{BH}^2C_{BH}}\sim -10^{-38}GeV^{-2}~.
\eeq

{\bf Dark Matter:} We infer the dark matter temperature by equating
the dark matter kinetic energy with thermal energy: 
\beq
T_{DM}\sim\left(v\over 100km/s\right)^2{m_{DM}\over 100GeV}10^{-4}GeV
\sim 10^{-4}GeV~,
\eeq 
with $m_{DM}$ being the mass of the dark matter particle. We consider
that only a fraction of the energy differential $dE$, of order  
$v^2/c^2\sim 10^{-3}$, goes 
into thermal energy.  These leads to a dark matter heat capacity of order
\beq\label{cdm}
C_{DM}\sim \pm 10^{-6}~,~~~~~~~{1\over T^2_{DM}C_{DM}}\sim  \pm 10^{-2}GeV^{-2}~.
\eeq 
In virialized bound systems there would be a negative  
contribution coming from the gravitational energy twice as large as
the kinetic component leading to a negative heat capacity (and the
negative sign in Eqn. \ref{cdm}). Non-bound dark matter would
contribute with a positive sign. We allow both signs in Eqn. \ref{cdm}
because our analysis is not detailed enough to consider which effect
dominates. 

{\bf Dark Energy:} We use the de Sitter entropy 
$S_{DE}=E^2/m_{Pl}^2\sim 10^{120}$ 
(with $E\equiv
\rho_{DE}H^{-3}$) giving 
\begin{equation}
d^2S_{DE}/dE^2\sim 2/m_{Pl}^2\sim 10^{-40}GeV^{-2}  
\label{eqn:SdS}
\end{equation}
  with a temperature of order $T_{DE}\sim H_0 \sim 10^{-42}GeV$. 


{\bf Total for the Universe:} Because the Universe is comprised of
different components which are not in equilibrium, we work with
Eqn. \ref{t2c} which is easy to treat as a sum of independent
components. 
Plugging all four components into Eqn. \ref{t2c} (with
$i=\{R,~BH,~DM,~DE\}$) leads to an expression of the form
\beq
{1\over T^2C}=-\sum_i{d^2S_{i}\over dE^2}~,
\eeq
to be compared with $(1+\beta Q)dS/(E_0-E_s)dE$ (from Eqn. \ref{t2cR}) for the 
random Hamiltonian. 

We notice that the ratios 
of $S_i$ to $dS_i/dE=T_i^{-1}$ and of $dS_i/dE$ to $d^2S_i/dE^2=T^{-2}_iC^{-1}_i$ 
for each component in the above estimates are all of order 
$E_0$. The regularity of these ratios makes it possible to reconcile 
the two descriptions if the following relation holds:
\begin{equation}\label{es}
\frac{1+\beta Q}{E_S-E_0}\sim\frac{1}{E_0}~,
\end{equation}  
which at this point of our analysis does not lead to any inconsistency with our
 previous 
results since the parameter $E_S$ was still unconstrained.  

Indeed, the generalized distributions we proposed, Eqns. (\ref{dr}) and 
(\ref{df}), have more free parameters than constraining equations, 
Eqns.(\ref{m1})-(\ref{delta2}), even after setting
 $\alpha=1/2$. Therefore, it appears that demanding consistency as we
have done above does not produce onerous constrains on the system.
A caveat to this conclusion could come from any insights that suggest
 that the properties of ratios of derivatives scaling as $E_0^{-1}$ is
 non-trivial for the actual Universe, but on the face of it this seems to be a
 straightforward result that obtains for a great variety of functional
 forms for  $S(E)$. 

An interesting feature of the above discussion is that it applies to a
variety of different cases:  The entropy and its various derivatives calculated
above are clearly dominated by the contributions from $S_\Lambda$.
But one could ``conservatively'' argue that $S_\Lambda$ is quite abstractly defined, and
should not be allowed to contribute to comparisons with the Wigner
density of states.  Perhaps the Wigner density of states should only be
equated with degrees of freedom that are more physically observable.  Removing
$S_\Lambda$ from the computation would allow $S_{BH}$ to dominate.
Since ratios of derivatives of $S_{BH}$ have the same properties, the
comparison with Wigner goes through unchanged.  Similar arguments
might cause one to leave out $S_{BH}$ as well.  Then $S_{DM}$
dominates and again the analysis goes through.  

Interestingly, if one considers the dark matter to be dominant, one
can consider integrating the discussion here with the comparison of
Wigner with field theory in Minkowski space in Sections
\ref{sec:FTWigBasic} and \ref{sec:FTWigDim}. The possibility that most
of the dark matter entropy is in states that are only linearly
perturbed gravitationally is consistent with current observations, and
under those conditions it may be reasonable to {\em combine} the
constraints presented here with those from Section \ref{sec:FTWig}.  
The value of $E_S$ needed to satisfy Eqn. (\ref{es}) together with the
field theory requirements is exponentially close to $-E_M$, half the
width of the Wigner distribution, with $E_0\ll E_M$. 

\vspace{.1in}

What are we to make of this comparison?  We are trying to learn if the
Wigner semicircle gives a sufficiently good approximation to the
density of states of the Universe.  Our current analysis assumes that
it is possible to take the Wigner semicircle density of states in the vicinity 
of some energy $E_0$ and set up a correspondence with eigenstates of a
Hamiltonian that describes the Universe more or less as we know it.  
In this section we assume this correspondence allows us to use the 
thermodynamic quantities as estimated above. Specifically, the differentiation 
with respect to $E$ should reflect the differences between the
thermodynamic quantities calculated at $E_0$, and for a similar
cosmological interpretation of the Wigner density of states an energy
$dE$ away.  A careful understanding of how the black holes, radiation,
etc. change as one shifts by $dE$ and reinterprets the density of
states cosmologically would be required to give our calculations more
rigor (of the sort commonly expressed, for example, by holding
specified properties fixed when differentiating thermodynamic
quantities).  In the absence of such rigor, we hope that the simple
differentiations preformed in this section give a reasonable
approximation to the desired result. 

The crudeness of our methods warrant a great deal of caution, but we
still find it a curiosity, perhaps even an encouraging curiosity that
our comparison yields results that are comparable within an order of
magnitude, and possibly even with the right sign.  


\section{Summary and Conclusions}
\label{sec:Conclusions}
The clock ambiguity suggest that we must view physical laws as
emergent from a random ensemble of all possible laws. We started this article
with a review of our earlier work showing the origin of the clock ambiguity.
We then outlined and expanded upon our earlier ideas about the
central role of quasiseparability in such a statistical analysis, and
discussed how this could lead to a prediction that local field theory should
provide the basic form for physical laws.  We have shown that
(contrary our earlier assumptions) one can quite generally assume
physical laws that emerge in this picture will have a time independent
Hamiltonian. We reviewed our earlier work that shows how the density of
states of a free field theory can be well approximated by a random
Hamiltonian, and extended this work to include a possible predictive
link to the number of dimensions of space. We also explored a higher
order analysis that (favorably) compares the curvature of the density
of states of a random Hamiltonian with that of the observed Universe
using estimates of the specific heat of the various components of the
Universe. 

While most of our discussion here is rather heuristic, our new results all add
to the case that a statistical approach to physical laws may indeed be
viable.  In the case of the time independence of the Hamiltonian, we
feel we have presented a very solid result which gives a significant
improvement over our earlier discussions. All in all, while many open
questions remain that could ultimately undermine our approach, we feel
that a statistical approach to the emergence of physical laws remains
an interesting possibility which has accumulated additional support
from the work presented here.

\begin{theacknowledgments}
  We would like to thank T.~Banks, S.~Deser, R.~Emparan, B.~Fiol, J.~Garriga,
  L.~Knox, H.~Nielsen,   L.~Smolin, G.~Starkman, A.~Tyson and S. White for 
valuable discussions. We also thank the organizers for an excellent
  meeting.  This work was supported in part by DOE Grant 
DE-FG03-91ER40674  

\end{theacknowledgments}



\bibliographystyle{aipproc}   

\bibliography{NYAS}


\end{document}